\documentclass{elsart}

 \usepackage{graphicx}
 \usepackage{epsfig}

\usepackage{amssymb}

\begin{document}

\begin{frontmatter}

\title{
Predictions of $pp,\bar pp$ total cross section and\\
$\rho$ ratio at LHC and cosmic-ray energies
}

\author{Keiji Igi$^a$ and Muneyuki Ishida$^b$}
\address{$^a$Theoretical Physics Laboratory, RIKEN, Wako, Saitama 351-0198, Japan}
\address{$^b$Department of Physics, Meisei University, Hino, Tokyo 191-8506, Japan}

\begin{abstract}
We propose to use rich informations on the $pp,\bar pp$ total cross sections $\sigma_{\rm tot}$  
below $N(\sim 10)$GeV in order to predict the total cross section and $\rho$ ratio 
at very high energies. 
Using the FESR as a constraint for high energy parameters,
we search for the simultaneous best fit to the data points of 
$\sigma_{\rm tot}$ and $\rho$ ratio up to some energy (e.g., ISR, Tevatron)
to determine the high-energy parameters.
We then predict $\sigma_{\rm tot}$ and $\rho$ in the LHC and high-energy cosmic-ray regions.
Using the data up to $\sqrt s=1.8$TeV(Tevatron), we predict
$\sigma_{\rm tot}^{pp}$ and $\rho^{pp}$ at the LHC energy($\sqrt s=14$TeV)
as $106.3\pm 5.1_{\rm syst} \pm 2.4_{\rm stat}$mb and 
$0.126\pm 0.007_{\rm syst}\pm 0.004_{\rm stat}$, respectively.
The predicted values of $\sigma_{\rm tot}$ in terms of the same parameters
are in good agreement 
with the cosmic-ray experimental data sample up to $P_{lab}\sim 10^{8\sim 9}$GeV
by Block, Halzen, and Stanov.
\end{abstract}

\end{frontmatter}


   Recently\cite{[1]}, we have proposed to use rich informations on $\pi p$ total cross sections 
below N($\sim$10 GeV) in addition to high-energy data to discriminate 
whether these cross sections increase like log $\nu$  or log$^2\ \nu$  at high energies\cite{[2]}. 
The FESR which was derived in the spirit of the $P^\prime$ sum rule\cite{[3]}
as well as the $n=1$ moment FESR(\cite{[4]}, \cite{[5]}) have been required to constrain 
the high-energy parameters. We then searched for the best fit of $\sigma_{\rm tot}^{(+)}$ above 70GeV 
in terms of high-energy parameters constrained by the two FESR.  
We then arrived at the conclusion that our analysis prefers the log$^2\nu$  
behaviours consistent with the Froissart-Martin unitarity bound\cite{[6]}.

   As for the $\bar pp$ and $pp$ total cross sections, there are a lot of data 
including cosmic-ray data 
up to $\sqrt s \sim$ several times of $10^4$GeV compared with data up to $\sqrt s \sim$30GeV 
for $\pi N$ scattering. 
Therefore, it is very valuable if one could investigate the high-energy behaviours at 
LHC and cosmic-ray regions\cite{Cosmic} using the similar approach as ref.~\cite{[1]}.

\begin{figure}
\includegraphics{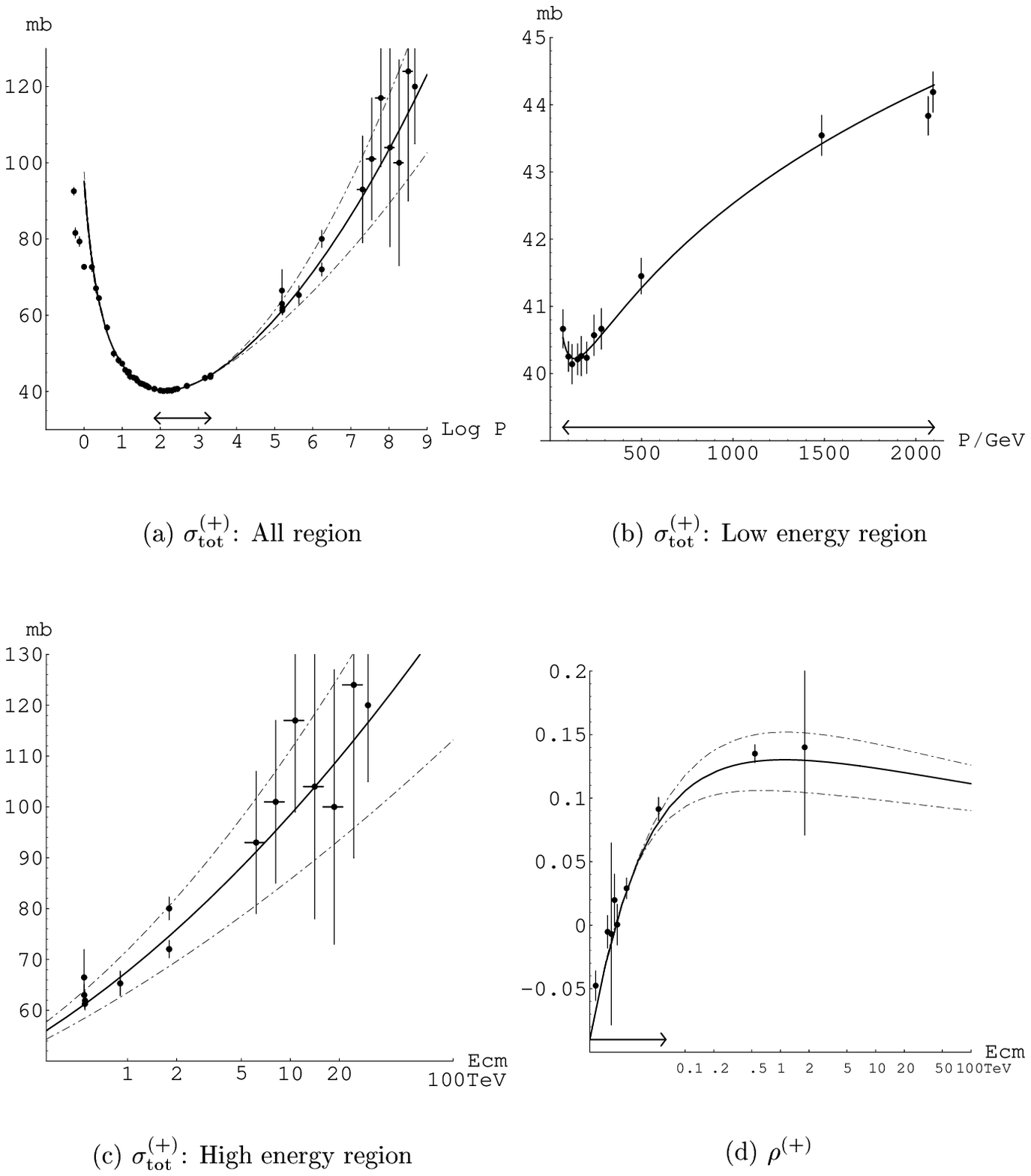}
\caption{\label{fig:1} Predictions for $\sigma^{(+)}$ and $\rho^{(+)}$ in terms of the fit 1.
The fit is done for the data up to the ISR energy, in the region 70GeV$\le$ $k$ $\le$ 2100GeV
(11.5GeV $\le \sqrt s \le$ 62.7GeV) which is shown by the arrow in each figure. 
Total cross section $\sigma^{(+)}_{\rm tot}$} in 
(a) all energy region, versus log$_{10}P_{lab}/$GeV,
(b) low energy region (up to ISR energy), versus $P_{lab}/$GeV and
(c) high energy (Tevatron-collider, LHC and cosmic-ray energy) region, 
    versus center of mass energy $E_{cm}$ in TeV unit.
(d) gives the $\rho^{(+)}(=Re\ F^{(+)}/Im\ F^{(+)})$ in high energy region, 
    versus $E_{cm}$ in terms of TeV. 
The thin dot-dashed lines represent the one standard deviation.  
\end{figure}

\begin{figure}
\includegraphics{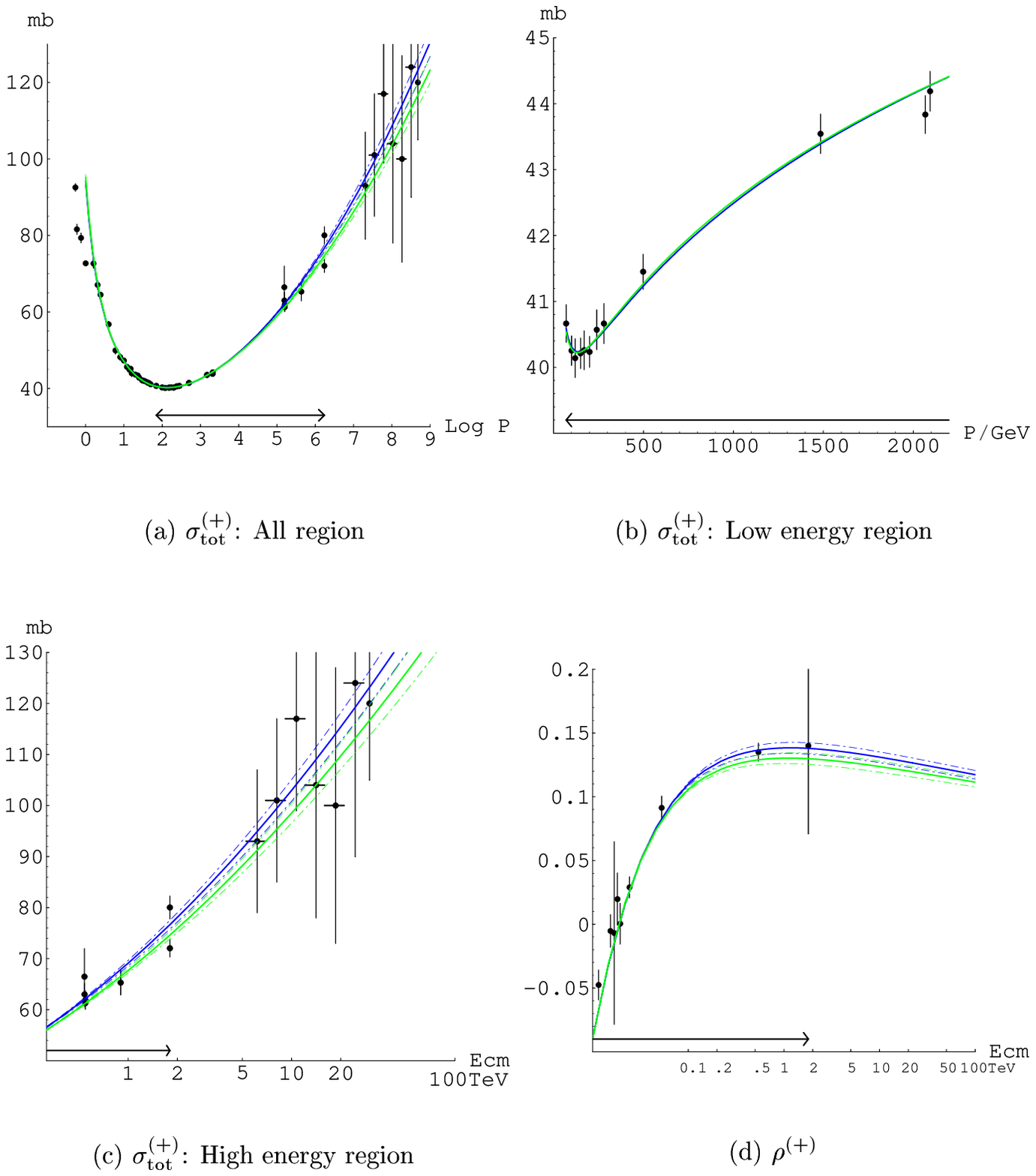}
\caption{\label{fig:2} Predictions for $\sigma^{(+)}$ and $\rho^{(+)}$ in terms of 
the fit 2(shown by green lines) and fit 3(shown by blue lines).
The fit is done for the data up to Tevatron-collider energy, in the region  
70GeV$\le$ $k$ $\le$ $2\times10^6$GeV(11.5GeV $\le \sqrt s \le$ 1.8TeV) 
which is shown by the arrow. 
For $k=1.7266\times 10^6$GeV($\sqrt s=E_{\rm cm}=1.8$TeV), 
the averaged datum of E710\cite{12d}/E811\cite{12c}, $\sigma_{\rm tot}^{\bar pp}=72.0\pm 1.7$mb,
is used in fit 2, while the $\sigma_{\rm tot}^{\bar pp}=80.03\pm 2.24$mb of CDF\cite{12b} is used
in fit 3.
For each figure, see the caption 
in Fig.\ref{fig:1}.    }
\end{figure}

(\underline{The purpose of this Letter}):\ \ \  
The purpose of this Letter is to predict $\sigma_{\rm tot}^{(+)}$, the $\bar pp$, $pp$ total cross 
sections and $\rho^{(+)}$, the ratio of the real to imaginary part of the forward scattering amplitude 
at the LHC and the higher-energy cosmic-ray regions, using the experimental data for 
$\sigma_{\rm tot}^{(+)}$ and $\rho^{(+)}$ for 70GeV$<P_{lab}<P_{large}$ as inputs. 
We first choose $P_{large}=2100$GeV corresponding to ISR region($\sqrt{s}\simeq 60$GeV). 
Secondly we choose $P_{large}=2\times 10^6$GeV corresponding to the Tevatron collider 
($\sqrt{s}\simeq 2$TeV). 
In a recent paper, Block and Halzen\cite{[8]} emphasized the importance of $\rho$ for the evidence 
for saturation of the Froissart-Martin bound\cite{[6]}.
We also use the $\rho$ ratio as input data in addition to FESR as a constraint.  
We searched for the simultaneous best fit of $\sigma_{\rm tot}^{(+)}$ and $\rho^{(+)}$ 
in terms of high-energy parameters $c_0,c_1,c_2$ and $\beta_{P^\prime}$ 
constrained by the FESR. 
It turns out that the prediction of $\sigma_{\rm tot}^{(+)}$ agrees with $pp$ experimental data 
at these cosmic-ray energy regions\cite{Cosmic,[e]} within errors in the first case ( ISR ). 
It has to be noted that the energy range of predicted $\sigma_{\rm tot}^{(+)}$, $\rho^{(+)}$ 
is several orders of magnitude larger than the energy region of 
$\sigma_{\rm tot}^{(+)}$, $\rho^{(+)}$ input (see Fig.~\ref{fig:1}).  
If we use data up to Tevatron (the second case), 
the situation is much improved, although there are some systematic uncertainties
coming from the data at $\sqrt s=1.8$TeV (see Fig.~\ref{fig:2}).

  FESR(1): Firstly we derive the FESR in the spirit of the $P^\prime$  
sum rule [3]. Let us consider the crossing-even forward scattering amplitude defined by
\begin{eqnarray}
F^{(+)}(\nu ) &=& \frac{f^{\bar pp}(\nu )+f^{pp}(\nu )}{2}\ \    
{\rm with}\ \  Im\ F^{(+)}(\nu )=\frac{k\ \sigma^{(+)}_{\rm tot}(\nu )}{4\pi}\ .
\label{eq1}
\end{eqnarray}

We also assume 
\begin{eqnarray}
Im\ F^{(+)}(\nu ) &=& Im\ R(\nu )+ Im\ F_{P^\prime}(\nu )\nonumber\\
 &=& \frac{\nu}{M^2}\left( c_0 + c_1 {\rm log}\frac{\nu }{M} + c_2 {\rm log}^2\frac{\nu }{M}  \right)
  + \frac{\beta_{P^\prime}}{M}\left( \frac{\nu}{M} \right)^{\alpha_{P^\prime}}\ \ \ \ \ 
\label{eq2}
\end{eqnarray}
at high energies ($\nu > N$).  We have defined the functions $R(\nu )$ and $F_{P^\prime} (\nu )$ 
by replacing $\mu$ by M in Eq.~(3) of ref.\cite{[1]}.
Here, $M$ is the proton( anti-proton) mass and $\nu ,\ k$ are the incident proton(anti-proton) 
energy, momentum in the laboratory system, respectively.

Since the amplitude is crossing-even, we have
\begin{eqnarray}
R(\nu ) &=& \frac{i\nu}{2M^2}\left\{ 2c_0+c_2\pi^2 
  + c_1 \left({\rm log}\frac{e^{-i\pi}\nu }{M}+{\rm log}\frac{\nu}{M}\right) \right. \nonumber\\
&& \left.  + c_2 \left({\rm log}^2\frac{e^{-i\pi}\nu }{M} + {\rm log}^2\frac{\nu}{M}\right)  \right\}\ ,\ \ \ \ \ \ \ 
 \label{eq3}\\
F_{P^\prime}(\nu ) &=& -\frac{\beta_{P^\prime}}{M}
 \left( \frac{(e^{-i\pi}\nu /M)^{\alpha_{P^\prime}}
       +(\nu /M)^{\alpha_{P^\prime}}}{{\rm sin}\pi\alpha_{P^\prime}} \right),
\label{eq4}
\end{eqnarray}
and subsequently obtain 
\begin{eqnarray}
Re\ R(\nu ) &=& \frac{\pi\nu}{2M^2}\left( 
  c_1 + 2 c_2 {\rm log}\frac{\nu}{M} \right)\ ,\ \ \  
 \label{eq5}\\
Re\ F_{P^\prime}(\nu ) &=& -\frac{\beta_{P^\prime}}{M}
 \left( \frac{\nu}{M}\right)^{0.5}\ ,\ \ \ 
\label{eq6}
\end{eqnarray}
substituting  $\alpha_{P^\prime} =\frac{1}{2}$ in Eq.~(\ref{eq4}). Let us define
\begin{eqnarray}
\tilde F^{(+)}(\nu ) &=& F^{(+)}(\nu )-R(\nu )-F_{P^\prime}(\nu) \sim \nu^{\alpha (0)}
\ (\alpha (0)<0)\ .    
\label{eq7}
\end{eqnarray}
Using the similar technique to ref.\cite{[1]}, we obtain
\begin{eqnarray}
Re\ \tilde F^{(+)}(M) &=& \frac{2 P}{\pi} \int_0^\infty 
         \frac{\nu Im\ \tilde F^{(+)}(\nu )}{k^2} d\nu \ \nonumber\\    
 &=& \frac{2 P}{\pi} \int_0^M 
         \frac{\nu}{k^2} Im\ F^{(+)}(\nu ) d\nu 
    +\frac{1}{2\pi^2} \int_0^{\overline{N}} 
         \sigma_{\rm tot}^{(+)}(k) dk \ \nonumber\\    
 & & - \frac{2 P}{\pi} \int_0^N 
         \frac{\nu}{k^2} \left\{ Im\ R(\nu ) 
  + \frac{\beta_{P^\prime}}{M}\left(\frac{\nu}{M}\right)^{0.5}  \right\} d\nu \ , 
\label{eq8}
\end{eqnarray}
where $\overline{N}=\sqrt{N^2-M^2} \simeq N$.  Let us call Eq.~(\ref{eq8}) as the FESR(1).

(\underline{FESR(2)}): The second FESR corresponding to $n=1$ [5] is:
\begin{eqnarray}  &&
\int_0^M \nu Im\ F^{(+)}(\nu )d\nu 
     + \frac{1}{4\pi}\int_0^{\overline{N}} k^2 \sigma_{\rm tot}^{(+)}(k)dk \nonumber\\
 & &=  \int_0^N \nu Im\ R(\nu ) d\nu 
     + \int_0^N \nu Im\ F_{P^\prime}(\nu ) d\nu \ \ \ . \ \ \  
\label{eq9}
\end{eqnarray}
We call Eq.~(\ref{eq9}) as the FESR(2) which we use in our analysis.

(\underline{The  $\rho^{(+)}$ ratio}): The $\rho^{(+)}$ ratio, 
the ratio of the real to imaginary part of 
 $F^{(+)}(\nu )$ is obtained from Eqs.~(\ref{eq2}), (\ref{eq5}) and (\ref{eq6}) as
\begin{eqnarray}
\rho^{(+)}(\nu ) &=& \frac{Re\ F^{(+)}(\nu )}{Im\ F^{(+)}(\nu )}
  = \frac{Re\ R(\nu )+Re\ F_{P^\prime}(\nu )}{Im\ R(\nu )+Im\ F_{P^\prime}(\nu )} \nonumber\\
  &=& \frac{ \frac{\pi\nu}{2M^2}\left( c_1+2c_2 {\rm log} \frac{\nu}{M} \right) 
          -\frac{\beta_{P^\prime}}{M}\left(\frac{\nu}{M}\right)^{0.5} }{
                  \frac{k\sigma_{\rm tot}^{(+)}(\nu)}{4\pi} }\ .\ \ \ 
\label{eq10}
\end{eqnarray}

(\underline{General approach}):\ \ \ 
The FESR(1)(Eq.~(\ref{eq8})) has some problem. i.e., there are the so-called 
unphysical regions coming from boson poles below the $\bar pp$ threshold.
So, the contributions from unphysical regions of the first term of the right-hand side
of Eq.~(\ref{eq8}) have to be calculated.
Reliable estimates, however, are difficult. 
Therefore, we will not adopt the FESR(1).

On the other hand, contributions from the unphysical regions to the first term of the
left-hand side of FESR(2)(Eq.~(\ref{eq9})) can be estimated to be an order of
0.1\% compared 
with the second term.\footnote{The average of the imaginary part from boson
resonances below the $\bar pp$ threshold is the smooth extrapolation of the $t$-channel
$qq\bar q\bar q$ exchange contributions from high energy to $\nu\le M$ 
due to FESR duality\cite{[4],[5]}.
Since
$Im\ F^{(+)}_{qq\bar q\bar q}(\nu ) < Im\ F^{(+)}(\nu )$,
$\int_0^M \nu Im\ F^{(+)}_{qq\bar q\bar q}(\nu )d\nu 
< \int_0^M \nu Im\ F^{(+)}(\nu ) d\nu = \int_0^M \frac{\nu}{2} Im\ f^{\bar pp}(\nu ) d\nu
\simeq \frac{M^2}{4} \left. Im\ f^{\bar pp}\right|_{k=0}
\simeq 3.2{\rm GeV} \ll \frac{1}{4\pi} \int_0^{\overline{N}} k^2 \sigma_{\rm tot}^{(+)}(k)dk
=3403\pm 20$GeV, where we use the experimental value, 
$\frac{k}{4\pi}\sigma_{\rm tot}^{\bar pp}\simeq$14.4GeV$^{-1}$ in $k<\ 0.3$GeV.
So, resonance contributions to the first term of Eq.~(\ref{eq9}) is less than 0.1\% of 
the second term.

Besides boson resonances, there may be additional contributions from multi-pion contributions 
below $\bar pp$ threshold. In the $\bar pp$ annihilation, $\bar pp\rightarrow \pi\pi$ could 
give comparable contributions with $\rho$-meson, but multi-pion contributions are suppressed 
due to the phase volume effects. Therefore, the first term of Eq.~(\ref{eq9}) will still be 
negligible even if the above contributions are included.} 
Thus, it can easily be neglected.

Therefore, the FESR(2)(Eq.~(\ref{eq9})), 
 the formula of $\sigma_{\rm tot}^{(+)}$(Eqs.~(\ref{eq1}) and (\ref{eq2})) 
 and the $\rho^{(+)}$ ratio (Eq.~(\ref{eq10})) are our starting points.
Armed with the FESR(2), we express high-energy parameters
$c_0,c_1,c_2,\beta_{P^\prime}$ in terms of the integral of total cross sections up to
$N$. 
Using this FESR(2) as a constraint for $\beta_{P^\prime}=\beta_{P^\prime}(c_0,c_1,c_2)$,
the number of independent parameters is three.
We then search for the simultaneous best fit to the data points of $\sigma_{\rm tot}^{(+)}(k)$
and $\rho^{(+)}(k)$ for 70GeV$\le k \le P_{large}$ to determine the values of
$c_0,c_1,c_2$ giving the least $\chi^2$. 
We thus predict the $\sigma_{\rm tot}$ and $\rho^{(+)}$ 
in LHC energy and high-energy cosmic-ray regions.

(\underline{Data}):\ \ \ 
We use rich data\cite{[7]} of $\sigma^{\bar pp}$ and $\sigma^{pp}$ to evaluate the relevant 
integrals of cross sections appearing in FESR(2). 
We connect the each data point\footnote{We take the error $\Delta y$ for each data point
$y$ as $\Delta y=\sqrt{(\Delta y)_{\rm stat}^2 + (\Delta y)_{\rm syst}^2 }$.
When several data points, denoted $y_i$ with error $\Delta y_i\ (i=1,\cdots,n)$,
are listed at the same value of $k$, these points are replaced by $\bar y$ with
$\Delta \bar y$, given by $\bar y=[\sum_i y_i/(\Delta y_i)^2 ] / [\sum_i 1/(\Delta y_i)^2 ]$
and $\Delta \bar y = \sqrt{  1 / [\sum_i 1/(\Delta y_i)^2 ] }$. 
Then, the data points with $\Delta \bar y$ less than 3 mb are picked up.
As a result, we obtain the 255(124) points for 
$k^2 \sigma_{\rm tot}^{\bar pp}$($k^2 \sigma_{\rm tot}^{pp}$), giving the integrals
$(5.070\pm 0.034)\times 10^4(\ (3.482\pm 0.037)\times 10^4 )$GeV in the region 
$0\le k \le \overline{N}$ with $\overline{N}=10$GeV.  
}
of $k^2\sigma_{\rm tot}^{\bar pp}$ and $k^2\sigma_{\rm tot}^{pp}$ with the next point by a 
straight line in order, from $k=0$ to $k=\overline{N}$, and regard the area of 
this polygonal line graph as the relevant integral in the region $0\le k \le \overline{N}$.
The integral of $k^2 \sigma_{\rm tot}^{(+)}(k)$ is given 
by averaging those of $k^2 \sigma_{\rm tot}^{\bar pp}(k)$ and $k^2 \sigma_{\rm tot}^{pp}(k)$.
We have obtained 
\begin{eqnarray}
\frac{1}{4\pi} \int_0^{\overline{N}} k^2 \sigma_{\rm tot}^{(+)}(k) dk 
 & = & 3403\pm20\ {\rm GeV}.  
\label{eq11}
\end{eqnarray}
for $\overline{N}=10$GeV (which corresponds 
to $\sqrt s=E_{cm}=4.54$GeV).\footnote{
The laboratory momentum $P_{lab}$ are related to the CM energy squared $s$ by
$s=2M(M+\sqrt{M^2+P_{lab}^2})$ and equivalently $P_{lab}=\frac{s}{2M}\sqrt{1-4M^2/s}$.
Thus, at high energies $E_{cm}=\sqrt s\simeq \sqrt{2M P_{lab}}$.
} 
The error of the integral, which is from the error of each data point,
is very small (less than 1\%), and thus, we regard the central value as an exact one in the 
following analysis. 

When $\sigma^{\bar pp}_{\rm tot}$ and $\sigma^{pp}_{\rm tot}$ data points are listed 
at the same value of $k$, we make the $\sigma_{\rm tot}^{(+)}(k)$ data point by averaging
these values. 
Totally, 37 points are obtained in the energy region, $0.54{\rm GeV}\le k\le 2100$GeV.
The data point of maximum value $k=2094.03$GeV($\sqrt s=62.7$GeV) comes from ISR\cite{ISR}. 
There are 12 points in the $70$GeV$\le k\le 2100$GeV (11.5GeV $\le \sqrt s \le$ 62.7GeV). 
There are no data reported in the wide range of $2100{\rm GeV} \le k \le 1\times 10^5$GeV.  
There are 6 points\cite{TEVATRON,12a,12b,12c,12d} of 
$\sigma^{\bar pp}_{\rm tot}$ in the Tevatron-collider energy region, 
$1\times 10^5{\rm GeV} \le k \le 2\times 10^6$GeV. 

It is necessary to pay special attention to treat the data with the maximum 
$k=1.7266\times 10^6$GeV($\sqrt s=1.8$TeV) in this energy range,
which comes from the three experiments E710\cite{12d}$/$E811\cite{12c} and CDF\cite{12b}.
The former two experiments are mutually consistent and their averaged $\bar pp$ cross section
is $\sigma_{\rm tot}^{\bar pp}=72.0\pm 1.7$mb, which deviates from the result of 
CDF experiment $\sigma_{\rm tot}^{\bar pp}=80.03\pm 2.24$mb.

Again there are no data reported in the range $2\times 10^6{\rm GeV}\le k \le 2\times 10^7$GeV.  
There are 7 points of $\sigma^{pp}_{\rm tot}$ with somewhat large errors, 
reported in the cosmic-ray energy region, 
$2\times 10^7{\rm GeV} \le k \le 5\times 10^8$GeV (6TeV $\le \sqrt s \le$ 30TeV), coming from 
cosmic-ray experiments\cite{Cosmic,[e]}.
Totally we obtain 25 points of $\sigma^{(+)}_{\rm tot}$ in $k\ge$ 70GeV.
We have not included the cosmic-ray data in our analysis. Thus, 18 points of 
$\sigma^{(+)}_{\rm tot}$ are used in the analyses. 

The data of $\rho^{\bar pp}(k)(=Re\ f^{\bar pp}(k)/Im\ f^{\bar pp}(k))$
 and $\rho^{pp}(k)(=Re\ f^{pp}(k)/Im\ f^{pp}(k))$ are reported in ref.\cite{[7]}.
When both data points are listed at the same value of $k$, we can make 
the $\rho^{(+)}(k)(=Re\ F^{(+)}(k)/Im\ F^{(+)}(k))$ 
data point.\footnote{\label{ft3}
Here the values of $Im\ f^{\bar pp}(k)$ and $Im\ f^{pp}(k)$ 
at the relevant values of $k$ are determined through the formula given in ref.\cite{[7]},
$\sigma^{\bar pp/pp}=Z + B {\rm log}^2 (s/s_0) + Y_1 (s_1/s)^{\eta_1} \pm Y_2 (s_1/s)^{\eta_2}$
with $(Z,B,Y_1,Y_2)$=$( 35.45, 0.308, 42.53, 33.34)$mb, $(\sqrt{s_0},\sqrt{s_1})$=$(5.38,1)$GeV
and $(\eta_1,\eta_2)$=$(0.458,0.545)$.  
} 
We obtain 9 points of $\rho^{(+)}$ in the energy region, 
$70{\rm GeV}\le k\le 2100$GeV.\footnote{
Here only the data point of maximum $k=1479$GeV($\sqrt s=52.7$GeV) is obtained 
by combining the $\rho^{\bar pp}$ at $k=1473.46$GeV($\sqrt s=52.6$GeV) 
and $\rho^{pp}$ at $k=1484.69$GeV($\sqrt s=52.8$GeV),
reported in ref.\cite{rho}. The other 8 points are obtained by combining 
$\rho^{\bar pp}$ and $\rho^{pp}$ with the same values of $k$.
}
No data are reported in the range $2100{\rm GeV}\le k \le 1\times 10^5$GeV.
The two points of $\rho^{\bar pp}$ are reported in Tevatron-collider energy region,
$1\times 10^5{\rm GeV}\le k \le 2\times 10^6$GeV (
at $k=1.5597\times 10^5$GeV($\sqrt s=541$GeV)\cite{rho1} and 
$k=1.7266\times 10^6$GeV($\sqrt s=$1.8TeV)\cite{12d} ).
We regard these two points as the $\rho^{(+)}$ data. 
As a result, we obtain 
11 points of $\rho^{(+)}$ up to Tevatron-collider energy region, 
$70{\rm GeV}\le k \le 2\times 10^6$GeV.

In the actual analyses, 
we use $Re\ F^{(+)}$ instead of $\rho^{(+)}(=Re\ F^{(+)}/Im\ F^{(+)})$.
The data points of $Re\ F^{(+)}(k)$ are made by multiplying $\rho^{(+)}(k)$ by
$Im\ F^{(+)}(k)=\frac{k}{8\pi}(\sigma_{\rm tot}^{\bar pp}(k)+\sigma_{\rm tot}^{pp}(k))$.
The values of $\sigma^{\bar pp}_{\rm tot}$ and $\sigma^{pp}_{\rm tot}$
at the relevant values of $k$ are obtained as follows: 
For $k<1500$GeV, they are determined by the formula given 
in ref.\cite{[7]}(see the footnote \ref{ft3}).
Two experimental values\cite{12a,12d} of $\sigma^{\bar pp}$  
in the Tevatron region are used.

(\underline{Analysis}):\ \ \ 
As was explained in the general approach, 
both $\sigma_{\rm tot}^{(+)}$ and $Re\ F^{(+)}$ data in 70GeV $\le k \le P_{large}$ 
are fitted simultaneously through the formula Eq.~(\ref{eq2}) and Eq.~(\ref{eq10}) 
with the FESR(2)(Eq.~(\ref{eq9})) as a constraint. 
FESR(2) with Eq.~(\ref{eq11}) gives us
\begin{eqnarray}
8.87 &=& c_0 + 2.04 c_1 + 4.26 c_2 + 0.367 \beta_{P^\prime}\ ,
\label{eq12}
\end{eqnarray}
which is used as a constraint of $\beta_{P^\prime}=\beta_{P^\prime}(c_0,c_1,c_2)$,
and the fitting is done by three parameters $c_0,c_1$ and $c_2$.

We have done for the following three cases:\\
{\bf fit 1)}:\ \ \  The fit to the data up to ISR energy region,
       70GeV $\le k \le$ 2100GeV, 
       which includes 12 points of $\sigma_{\rm tot}^{(+)}$ 
       and 7 points of $\rho^{(+)}$. \\
{\bf fit 2)}:\ \ \  The fit to the data up to 
     Tevatron-collider energy region, 70GeV$\le k \le 2\times 10^6$GeV.
     For $k=1.7266\times 10^6$GeV($\sqrt s=1.8$TeV), the E710$/$E811 datum is used.
     There are 18 points of $\sigma_{\rm tot}^{(+)}$ 
       and 9 points of $\rho^{(+)}$.\\
{\bf fit 3)}:\ \ \  The same as fit 2, except for the CDF value at $\sqrt s=1.8$TeV used.

(\underline{Results of the fit}):\ \ \ 
The results are shown in Fig.~\ref{fig:1}(Fig.~\ref{fig:2}) for the fit 1(fit 2 and fit 3).
The $\chi^2/d.o.f$ are given in Table \ref{tab1}. 
%
%
The reduced $\chi^2$ and the respective $\chi^2$-values devided by the number of data points 
for $\sigma_{\rm tot}^{(+)}$ and $\rho^{(+)}$ are less than or equal to unity.  
The fits are successful in all cases.
There are some systematic differences between fit 2 and fit 3, which come from the
experimental uncertainty of the data at $\sqrt s=1.8$TeV mentioned above.

\begin{table}
\caption{ The values of $\chi^2$ for the fit 1 (fit up to ISR energy) and 
         the fit 2 and fit 3 (fits up to Tevatron-collider energy). 
$N_F$ and $N_\sigma (N_\rho )$ are the degree of freedom and 
the number of $\sigma^{(+)}_{\rm tot}(\rho^{(+)})$ data points in the fitted energy region.
}
\begin{tabular}{c|ccc|}
         & $\chi^2/N_F$        
            &  $\chi^2_{\sigma}/N_{\sigma}$ & $\chi^2_{\rho}/N_{\rho}$ 
  \\
\hline
fit 1  & 10.6/15 & 3.6/12 & 7.0/7\\  
fit 2  & 16.5/23 & 8.1/18 & 8.4/9\\
fit 3  & 15.9/23 & 9.0/18 & 6.9/9\\
\hline
\end{tabular}
\label{tab1}
\end{table}

The best-fit values of the parameters are given in Table \ref{tab2}.
Here the errors of one standard deviation are also 
given.\footnote{The $c_2$log$^2(\nu /M)$-term in Eq.~(\ref{eq2}) is most relevant for predicting 
$\sigma_{\rm tot}^{(+)}$ in high energy region, and thus,
the error estimation is done as follows: The $c_2$ is fixed with 
a value deviated a little from the best-fit value, and then the $\chi^2$-fit 
is done by two parameters $c_0$ and $c_1$. When the resulting $\chi^2$ is larger than 
the least $\chi^2$ of the three-parameter fit by one, the corresponding values of parameters
give one standard deviation. 
}

\begin{table}
\caption{
The best-fit values of parameters in the fit 1, fit 2 and fit 3.
}
\begin{tabular}{c|cccc|}
       & $c_2$  & $c_1$  & $c_0$  &  $\beta_{P^\prime}$ \\
\hline
fit 1  & $0.0411\pm 0.0199$ & $-0.074\mp 0.287$ & $5.92\pm 1.07$ & $7.96\mp 1.55$ \\
fit 2  & $0.0412\pm 0.0041$ & $-0.076\mp 0.069$ & $5.93\pm 0.28$ & $7.95\mp 0.44$ \\
fit 3  & $0.0484\pm 0.0043$ & $-0.181\mp 0.071$ & $6.33\pm 0.29$ & $7.37\mp 0.45$ \\
\hline
\end{tabular}
\label{tab2}
\end{table}

(\underline{Predictions for $\sigma^{(+)}$ and $\rho^{(+)}$ 
at LHC and cosmic-ray energy region}):\ \ \ 
By using the values of parameters in Table~\ref{tab2},
we can predict the $\sigma_{\rm tot}^{(+)}$ and $\rho^{(+)}$ in higher energy region, 
as are shown, respectively in (c) and (d) of Fig.~\ref{fig:1} and \ref{fig:2}. 
The thin dot-dashed lines represent the one standard deviation.

As is seen in (c) and (d) of Fig.~\ref{fig:1}, 
the fit 1 leads to
the prediction of $\sigma_{\rm tot}^{(+)}$ and $\rho^{(+)}$ with somewhat large errors in the 
Tevatron-collider energy region, although the best-fit curves are consistent 
with the present experimental data in this region. Furthermore, 
the predicted values of $\sigma_{\rm tot}^{(+)}$ agree with $pp$ experimental data 
at the cosmic-ray energy regions\cite{Cosmic,[e]} within errors (see (a),(c) of Fig.~\ref{fig:1}).
The best-fit curve gives $\chi^2/$(number of data) to be 13.0/16, 
and the prediction is successful.
As was mentioned in the purpose of this Letter,
it has to be noted that the energy range of predicted $\sigma_{\rm tot}^{(+)}$
is several orders of magnitude larger than the energy region of 
the $\sigma_{\rm tot}^{(+)}$, $\rho^{(+)}$ input.  
If we use data up to Tevatron-collider energy region as in the fit 2 and fit 3, 
the situation is much improved (see (a),(c) of Fig.~\ref{fig:2}),
although there is systematic uncertainty depending on the treatment 
of the data at $\sqrt s=1.8$TeV.

The best-fit curve gives $\chi^2/$(number of data) from cosmic-ray data, 1.3/7(1.0/7)
for fit 2(fit 3).

We can predict the values of $\sigma_{\rm tot}^{(+)}$ and $\rho^{(+)}$
at LHC energy, $\sqrt s$=$E_{cm}$=14TeV and 
at very high energy of cosmic-ray region.
The relevant energies are very high, and
the $\sigma_{\rm tot}^{(+)}$ and $\rho^{(+)}$ can be regarded to be equal to the
$\sigma_{\rm tot}^{pp}$ and $\rho^{pp}$.
The results are shown in Table~\ref{tab3}.

\begin{table}
\caption{
The predictions of $\sigma_{\rm tot}^{(+)}$ and $\rho^{(+)}$ 
at LHC energy $\sqrt{s}=E_{cm}=14$TeV($P_{lab}$=1.04$\times 10^8$GeV), and 
at a very high energy $P_{lab}=5\cdot 10^{20}$eV
($\sqrt s$=$E_{cm}$=967TeV.) 
in cosmic-ray region.
}
\begin{tabular}{c|cc|cc|}
        &  $\sigma_{\rm tot}^{(+)}$({\scriptsize $\sqrt s$=14TeV}) 
        &  $\rho^{(+)}$({\scriptsize $\sqrt s$=14TeV})
        &  $\sigma_{\rm tot}^{(+)}$({\scriptsize $P_{lab}$=$5\cdot 10^{20}$eV}) 
        &  $\rho^{(+)}$({\scriptsize $P_{lab}$=$5\cdot 10^{20}$eV})\\
\hline
fit 1   & $103.8\pm 14.3$mb & $0.122\stackrel{+0.018}{\scriptstyle -0.024}$
        & $188\pm 43$mb & $0.099\stackrel{+0.011}{\scriptstyle -0.017}$\\
fit 2   & $103.8\pm 2.3$mb  & $0.122\pm 0.004$
        & $189\pm 8$mb  & $0.100\pm 0.003$\\
fit 3   & $108.9\pm 2.4$mb  & $0.129\pm 0.004$
        & $204\pm 8$mb  & $0.104\pm 0.003$\\
\hline
\end{tabular}
\label{tab3}
\end{table}

The prediction by the fit 1 
in which data up to the ISR energy are used as input has somewhat large(fairly large) errors 
at LHC energy(at high energy of cosmic ray). By including the data up to the Tevatron collider,
the prediction of fit 2(using E710/E811 datum) is smaller than that of fit 3(using CDF datum).
We regard the difference between the results of fit 2 and fit 3 as the systematic uncertainties
of our predictions. As a result,
we predict 
\begin{eqnarray}
\sigma_{\rm tot}^{pp} &=&  106.3\pm 5.1_{\rm syst} \pm 2.4_{\rm stat}\ {\rm mb},\ \   
\rho^{pp} = 0.126\pm 0.007_{\rm syst}\pm 0.004_{\rm stat}\ \ \ \ \ \ \ \ 
\label{eq13}
\end{eqnarray}
at LHC energy($\sqrt s=E_{cm}=14$TeV).
We obtain fairly large systematic errors coming from the experimental unceratinty 
at $\sqrt s=1.8$ TeV.

The predicted central value of $\sigma_{\rm tot}^{pp}$ is in good agreement with 
Block and Halzen\cite{[a]} 
$\sigma_{\rm tot}^{pp}=107.4\pm 1.2$ mb, $\rho^{pp}=0.132\pm 0.001$. 
In contrary  to our results( see Fig.~2(a), (c)), however, their values are not affected 
so much about CDF, E710/E811 discrepancy. Our prediction has also to be compared with 
Cudell et al.\cite{[f]} 
$\sigma_{\rm tot}^{pp}=111.5\pm 1.2_{\rm syst}\stackrel{+4.1}{\scriptstyle -2.1}_{\rm stat}$ mb, 
$\rho^{pp}=0.1361\pm 0.0015_{\rm syst}\stackrel{+0.0058}{\scriptstyle -0.0025}_{\rm stat}$, 
who's fitting techniques favours the CDF point at $\sqrt s=1.8$ TeV.

Finally we emphasize that precise measurements of both $\sigma_{\rm tot}^{pp}$ and 
$\rho^{pp}$ in the coming LHC experiments will resolve the FNAL discrepancy of 
$\sigma_{\rm tot}^{pp}$ ( Fig.~2(a), (c)). The LHC measurements would also clarify 
which is the best solution among the three high-energy cosmic-ray samples\cite{[c],[d],[e]}.\\
{\it Note added in proof}:
After completion of the hep-ph/0505058, we were informed that M.~M.~Block and F.~Halzen\cite{[a]} 
have also done the similar work based on the same spirit of duality using different method 
independently. We were also informed by  M.~J.~Menon\cite{[b]} about other cosmic-ray analyses 
by Gaisser et al.\cite{[c]} and N.~N.~Nikolaev\cite{[d]} besides M.~M.~Block et al.\cite{[e]} 
which are used as cosmic-ray data in this Letter.



\end{document}